# Half or full core-hole in Density Functional Theory X-ray absorption spectrum calculations of water?

Matteo Cavalleri,[a] Michael Odelius,[a] Dennis Nordlund,[ab] Anders Nilsson,[ab] and Lars G.M. Pettersson[a*]

[a] *Fysikum - Stockholm University, AlbaNova University Center, S-10691 Stockholm, Sweden.*
[b] *Stanford Synchrotron Radiation Laboratory, P.O.B. 20450, Stanford, CA 94309, U.S.A.*
[*] *Fax: +46 8 5537 8442; Tel: +46 8 5537 8712; E-mail: lgm@physto.se*

**We analyze the performance of two different core-hole potentials in the theoretical modeling of XAS of ice, liquid and gas phase water; the use of a full core-hole (FCH) in the calculations, as suggested by Hetenyi et al. [J. Chem. Phys. 120 (18), 8632 (2004)], gives poor agreement with experiment in terms of intensity distribution as well as transition energies, while the half core hole (HCH) potential, in the case of water, provides a better compromise between initial and final state effects, leading to good agreement with the experimental data.**

It was recently suggested, based on x-ray absorption (XAS) and x-ray Raman spectroscopy (XRS) combined with Density Functional Theory (DFT) calculations of spectra, that water in the liquid is in an asymmetric hydrogen bonding situation with only two strong H-bonds: one donating and one accepting with the remaining bonds weakened or broken through mainly bending off the H-bond angle[1]. This provocative result has attracted much attention and debate in the literature. Smith et al.[2] analyzed the temperature dependence of XAS spectra obtained on a liquid microjet and claimed agreement with four-coordinated water. Their experimental data has, however, been shown to be flawed by saturation effects which invalidate their conclusions[3]. On the theory side Hetenyi et al.[4] computed spectra directly from their ab initio molecular dynamics (AIMD) simulations and claimed agreement with experiment by using a full core hole (FCH) in their spectrum calculations rather than the standard half core hole (HCH). In view of the predominance (80% at room temperature) of single-donor (SD) species reported in ref.[1] this is a surprising and contradictory result with respect to the XAS conclusions since the AIMD simulation only contained 19% SD species being instead dominated by double-donor (DD), ice-like coordination. In view of the present debate on the structure of water it is important to analyze in depth the validity of the HCH and FCH approximations to the core-hole potential in DFT calculations of XAS spectra.

XAS (elsewhere also NEXAFS or XANES)[5] and the corresponding non-resonant x-ray Raman spectroscopy (XRS)[6] are well-established experimental techniques that have recently been extended to measurements of the local electronic structure of liquid water[1,7-11]. Both XAS and XRS show major differences in the spectral profiles between bulk ice and liquid water[1,8,9]; there is a well-defined pre-edge structure that can be seen in the spectrum of the liquid and is substantially reduced in the bulk ice spectrum; the remaining intensity is mainly due to proton disorder and minor defects[12]. The liquid water spectrum, with strong pre-edge (535 eV) and main-edge (537 eV) features, instead closely resembles that of the ice surface where a dominant fraction of the water molecules in the first half-bilayer has one free O-H bond[1,13]. The spectrum of bulk ice, in which each molecule is tetrahedrally coordinated, is instead characterized by weak main edge (537 eV) and strong post edge (540 eV) features[1,8]; these experimental differences already indicate that the liquid cannot be dominated by fully coordinated molecules.

In ref.[1] density functional theory (DFT) based spectrum calculations using the standard transition potential or HCH approximation were used to interpolate spectra between the two experimental extremes, i.e. bulk and surface of ice. This confirmed the experimental analysis in terms of coordination and furthermore lead to an operational H-bond definition relative to a geometrical cone at each of the donor hydrogens which gives the distortion needed to give the pre- and main-edge features and which was used to evaluate MD simulation models. Considering the importance of the theoretical models for the analysis and the results for the full core-hole potential as used by Hetenyi et al.[4] we must critically evaluate the approach taken in ref.[1] to compute theoretical XAS spectra in this context.

In a one-electron picture the XA spectrum reflects the local unoccupied density of states[†]. According to the final state rule the energy positions of the spectral features reflect the eigenstates of the final state Hamiltonian, which includes a core-hole in the XA spectroscopy[14,15]. On the other hand, according to the initial state rule, the integrated x-ray transition intensity in the valence region should reflect the number of empty p-valence states in the initial state, *i.e.* before the x-ray transition takes place, which for XAS corresponds to the ground state[15]. In other words, in XA spectra the energy positions of the electronic levels are governed by the presence of the core hole, while the intensity distribution should correspond to that of the empty states with p-symmetry in the ground state, projected on the excited atom. Computing XAS spectra thus becomes a question of finding the correct balance between initial and final state contributions. One way to estimate the effect of the final state is to use the Z+1 approximation, which implies that the core-excited atom is replaced by the Z+1 atom. In XA spectrum calculations this full core-hole, final-state-dominated potential is obtained by removing a core electron (in all-electron calculations) or by specially designed pseudopotentials with occupation $1s^1$ (in *e.g* plane-waves-based spectrum calculations)[4,9,16]. A more well-established technique is the transition potential approach by Slater in which excitation energies are obtained as orbital energy differences between initial and final orbitals with half an electron excited[17]; in the context of XA spectrum calculations this implies that half a core-electron is excited in the calculations[17-19]. For excitation energies obtained this way it can be shown that relaxation effects are included up to second order[17]. Previous applications of the HCH approach have shown excellent agreement with experiment, e.g., in the case of solutions containing transition-metal ions[11] and water at interfaces[10], leading to a detailed interpretation of the data.

In the present communication we will systematically investigate the FCH and HCH potentials in DFT calculations of XA spectra for water in its different aggregation states[§]. The two approaches will be evaluated by comparison with experiment for gas phase, bulk and surface of ice, for which the structures are known, before we turn to a discussion of liquid water for which the structure is currently under debate[1,4]. We begin by discussing the

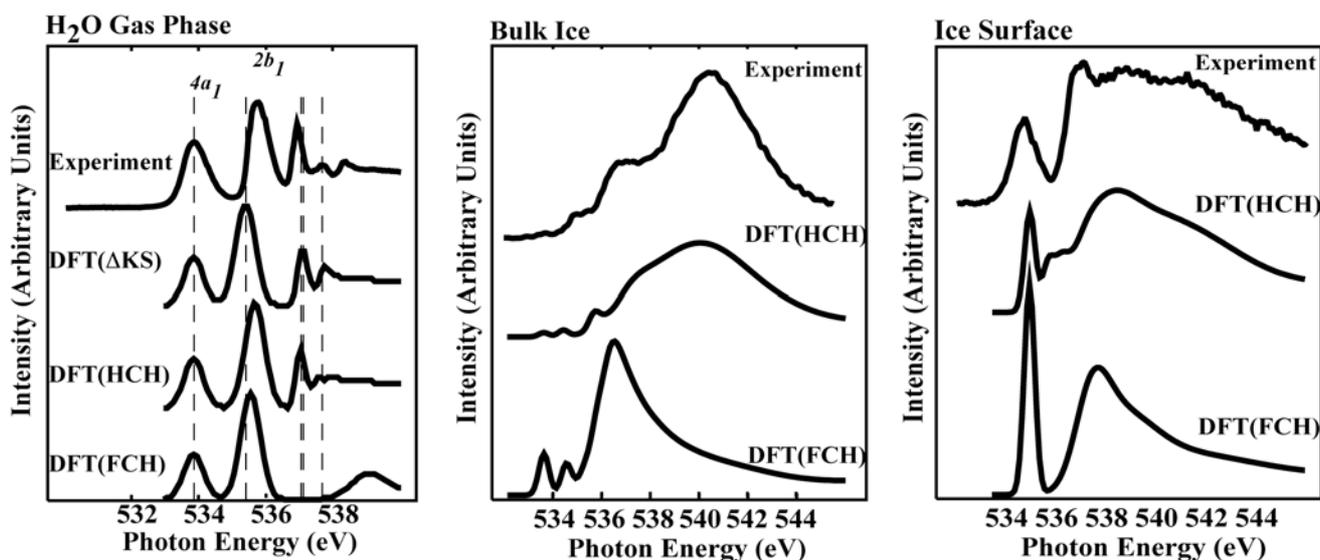

**Fig. 1** (left) Experimental XA spectrum of gas phase water from ref.[9] and computed O(1s) XA spectra for gas phase water molecules using DFT with half core-hole, DFT(HCH) and the full core-hole, DFT(FCH) methods. The ΔKohn-Sham approach is used to define the energy scale of the DFT-based calculations. Experimental XA spectrum from ref.[1] of bulk (middle) and ice surface (right), compared with the respective computed spectra using half, DFT(HCH), and full core hole, DFT(FCH). Vertical dashed lines represent the position of the variationally computed 5 lowest transition energies.

gas phase water XA spectra in Figure 1; the geometry was taken as the experimental gas phase geometry. The computed spectra were generated by a Gaussian convolution of the discrete oscillator strengths with varying broadenings corresponding to the widths of the features in the experimental spectrum. For the first two peaks, $4a_1$ and $2b_1$, that are strongly anti-bonding and have very broad Franck-Condon profiles in the experiment, the broadening (fwhm) was 0.7 eV, while for the sharp Rydberg series starting at higher energies a fwhm of 0.3 eV was used. The absolute energy scale of the computed DFT spectra was determined by shifting the spectra to coincide with the first variationally computed excitation energy[§]. The HCH spectrum was thus shifted downwards 2.5 eV while, in the case of the FCH calculation, the resulting downwards shift based on the same variational calculation of the first excited state is substantially larger, 13.7 eV. The balance between final and initial state effects, provided by the HCH approach, results in an excellent agreement of the simulated spectrum with experiment.

All peaks, including also the first Rydberg transitions, are consistent in intensity and position with what is seen experimentally and also compare extremely well with the DFT(ΔKS) spectrum. In the latter the five lowest transitions from the HCH calculation have been shifted to correspond to the variationally computed transition energies, displayed in fig. 1 as vertical dashed lines, which represents, in principle, the best theoretical spectrum within the Kohn-Sham DFT framework[20]; the small resulting changes in the peak positions compared to the HCH spectrum show that differential relaxation effects in the excited states are small for this molecule within the HCH potential. The FCH spectrum, on the other hand, completely fails to reproduce the high-lying Rydberg peaks. The full core-hole potential overestimates the final state effects by contracting the 2p orbitals and consequently screening the 3p and higher Rydberg states. This causes the gap between the $2b_1$ and the Rydberg sequence to increase in the FCH spectrum.

In bulk ice the water molecules are tetrahedrally coordinated, with each water molecule taking part in four H-bonds. The bulk ice XA spectrum differs from that of liquid water and ice surface[1] through an intensity shift from the pre (535 eV) and main (537 eV) edges to the post-edge (540 eV). In figure 1 we also compare the experimental bulk ice spectrum (Secondary electron yield, SEY XAS, from ref.[1]) with the computed spectra using the HCH and FCH methods. In both cases the tetrahedral ice (Ih) is modeled using a 17 water molecules cluster so that the excited molecule in the center of the cluster has both the first and second coordination shells completed. The half-core-hole calculation agrees quite well in general appearance with experiment; some minor discrepancies in the intensity ratio of the peaks are due to the limited size of the model[1,21]. The weak intensity of the pre-edge feature in the experimental ice spectrum can be related to the presence of defects and proton disorder in the crystal[12]. As in the case of the gas phase spectrum the FCH approximation artificially enhances localization of the core-excited states and brings down a state from the band structure resulting in a localized excitonic state, while higher-lying states are pushed up in energy. The computed FCH bulk ice spectrum is consequently characterized by the lack of intensity in the post-edge region (~540 eV) and is instead dominated by intensity at the main edge. Note that this is in serious disagreement with experiment both in terms of the overall shape of the spectrum as well as in terms of the localization.

Figure 1 (right, top) shows the XA spectrum of the topmost surface layer in an ice film; this spectrum was obtained by subtracting appropriately scaled normal emission from grazing emission Auger electron yield (AEY) spectra which allows to isolate the signature from molecules in the first half-bilayer characterized by one free O-H bond.[1,13] Upon adsorption of ammonia on the ice surface all relevant spectral features change, since the free O-H groups are now saturated[1], demonstrating that these features are indeed surface-related and result from structures accessible to H-bond formation with $NH_3$. The suggested existence of a dominant fraction of single-donor (SD) water molecules at the surface of ice is furthermore consistent with previous AEY and $H^+$ yield XAS studies[22]. In order to model the ice surface a 15 water molecules cluster was used in the spectrum calculations. The excited molecule is in a single-donor H-bonding conformation with one free O-H bond while the other is hydrogen bonded to the surface second bilayer. The theoretical spectra are qualitatively similar and in good agreement with the experimental data, although FCH overestimates the intensity of the pre-edge peak with respect to the main-edge. However, the general loss of intensity in the post edge region compared to experiment is evident also in this case.

For liquid water it was demonstrated by Wernet et al.[1] that the investigated molecular dynamics simulations, including *ab initio* Car-Parrinello molecular dynamics (CPMD)[23], all resulted in a preponderance of DD species with two donating H-bonds. XA spectra of DD molecules are characterized by the same features typical of the spectrum of bulk ice with weak pre- and main-edges and an intense and broad post-edge around 540 eV[1,8,9,21] and therefore a summed spectrum computed for a snapshot largely dominated by DD species cannot be expected to reproduce the spectral characteristics of the experimental liquid water spectrum, *i.e.* strong intensity in the pre- and main-edge regions. On the other hand, Hetényi and coworkers claim agreement with experiment from summing spectra computed using the FCH potential of all 64 molecules in a typical CPMD snapshot in spite of only 19% H-bonds being broken[4]. However, because of the lack of the 1s orbital in the pseudo-potential plane-waves approach in the CPMD framework it is not possible to obtain the absolute energy scale of the spectra and a shift of the excitation energies by an arbitrary[4,16] or a computed[21], based on all-electron calculations, constant is necessary to allow direct comparison with experiment.

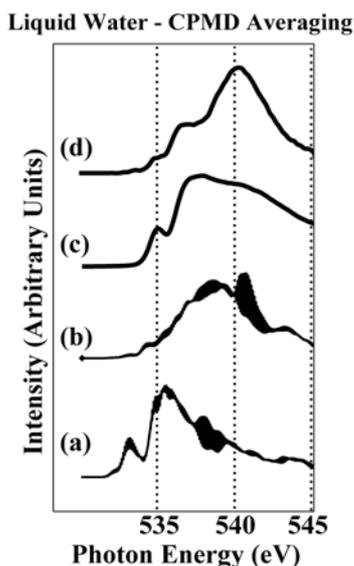

**Fig. 2.** Sums of computed spectra for all 32 water molecules in a representative CPMD snapshot using a) FCH, and b) HCH approximations. Experimental spectra of liquid water c), and bulk ice d), from ref.[1].

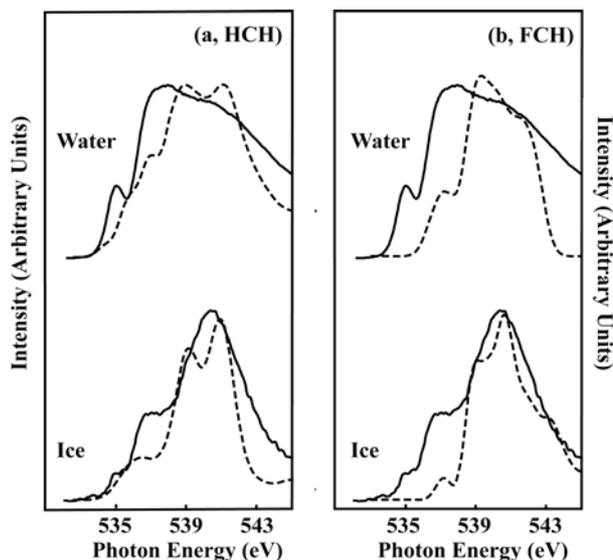

**Fig. 3.** Comparison between experiment from ref.[1] (solid lines) and the theoretical spectra from ref.[4] (dashed lines), obtained using a) the HCH and b) the FCH approximations. The theoretical spectra have been shifted upwards with respect to the original publication to be aligned with the post-edge region of the experimental bulk ice spectrum.

The use of the all-electron model-cluster approach is in this case advantageous since absolute excitation energies are available and the ΔKohn-Sham method makes it possible to compare HCH and FCH calculations without arbitrary energy shifts. For large enough unit cell in CPMD and cluster size in all-electron calculations it has been shown for water that, in the case of the HCH approximation, the computed spectra for different HB configurations using all-electron molecular clusters yield excellent agreement with those calculated within a periodic plane-waves DFT framework[21]. We may thus obtain spectra from periodic plane-waves CPMD simulations by using appropriate all-electron molecular clusters based on the CPMD structures. In figure 2 we show the summed calculated spectra for a typical CPMD snapshot (approximately 80% DD) using the FCH (2a) and HCH (2b) approximations within the all-electron STOBE-DEMON program. The final spectra were obtained by summing up the spectra of all 32 water molecules in the simulation box using a constant 0.5 eV broadening of all peaks. The calculations were performed by extracting clusters composed of 32 water molecules with the core-excited oxygen in the center of the simulation cell. The resulting spectrum using the HCH approximation (2b) is, as expected from the predominance of DD configurations, similar to the experimental spectrum of bulk ice (2d) and to the computed spectrum for the ice model shown in figure 1. The FCH spectrum (2a) has intensity shifted to lower energies and has the spectrum main-edge at 535 eV in the position of the experimental pre-edge peak of liquid water (2c); an additional peak appears at low (≈533 eV) excitation energy, where the HCH spectrum shows no intensity although also in this case there is a density of states in this region. Note that the HCH and FCH spectra are energy-calibrated relative to the same computed first core-excited state, i.e. the energy of the first transition is the same. The FCH spectrum is furthermore characterized by the loss of intensity at the post-edge due to the artificial localization and downward shift of the 2p relative to the extended band states. The strong intensity in the pre-and main-edge regions is an artifact due to enhancement of final state effects in the FCH approximation as is evident by comparing the FCH liquid (with ≈80% DD configurations) (2a) and the FCH ice (figure 1) spectra. The overall spectral shapes are very similar and neither of them shows appreciable intensity in the post-edge region in strong contrast with experiment.

This behaviour is also evident if the theoretical spectra of ref.[4] are compared with the experimental data in ref.[1]. Because of the lack in ref.[4] of an absolute energy scale the spectra were shifted to maximize the agreement with experiment for water. If a shift should be applied we argue that both HCH and FCH spectra should be shifted to align with the experimental spectrum of ice, for which the structure is known, rather than liquid water for which the structure is under debate. This is done in figure 3 using the same shift on both ice (100% DD) and the CPMD simulated liquid (≈80% DD). The HCH approximation (Fig 3a) is unable to reproduce the sharp pre-edge peak, due to SD species, of the experimental spectrum of liquid water and presents the intensity at ~540 eV which is characteristic of the predominance of DD species in the simulation. The computed HCH ice spectrum is brought into overall agreement with experiment although a discrepancy in the intensity ratio between main- and post-edge peaks is evident; some of the intensity in the experimental bulk ice main edge can, however, still be due to defects[1,12]. The splitting of the post-edge peak in the calculation by Hetényi et al. is possibly due to the limited Brillouin zone sampling used in the calculation. The amount of broadening to be applied to states beyond the edge is furthermore uncertain. The FCH spectrum of liquid water from ref.[4] (fig 3b) is clearly not representing the experiment after calibration against ice, but shows a spectral shape resembling our corresponding FCH spectrum computed with all-electron DFT shown in figure 2a. The most significant discrepancy is the lack of intensity at higher energies (in both fig 2a and 3b) characteristic of the lack of hybridization between 2p and higher states due to the artificial stabilization of the 2p levels; this deficiency in the spectral shape is independent of energy calibration. In the all-electron approach the absolute excitation energies are, however, available and the spectra can thus be placed correctly against the experiment revealing that the computed FCH main-edge actually corresponds in position to the experimental pre-edge peak and the post-edge is largely absent. It is clear that neither the spectrum of bulk ice nor that of liquid water is well represented in the FCH approximation.

The qualitative effects of the FCH approach can be understood in relation to the initial state rule by considering a nearly closed-shell electronegative atom such as oxygen or fluorine in a molecule. Upon core-excitation severe changes in the electron population are expected, due to the charge transfer from neighboring atoms. In the Z+1 approximation F ($2p^5$) is replaced by Ne, which has a completely filled electronic shell structure. The FCH approximation in this case will yield an XA spectrum with only Rydberg transitions to the 3p, 4p and the continuum, but not to the 2p states in serious discrepancy with experiment. Similarly, with oxygen being replaced by the very electronegative fluorine, the empty p-type holes will have an energy position according to the bonding energies of the empty levels for the F atom, but the intensity in the spectrum should rather follow the number of empty states of p-character of the oxygen atom. Because of its high electronegativity the F-core will induce charge-transfer from the neighboring atoms increasing its p population. This will have a huge impact on the spectral shape, especially on the intensities of lower-lying Rydberg peaks. For less electronegative elements, like C, these effects are less dramatic. In the Z+1 approximation C ($2p^2$) is replaced by N ($2p^3$); due to the smaller differences between initial and final state p population we can expect the FCH approximation to work better than in the case of O and F. Although in general there are pragmatic reasons for preferring Slater's transition potential or HCH approach, it has to be remembered that in different systems the balance of the influence from initial and final state effects on the spectra could vary, making the differences between the FCH and HCH potentials smaller than in the extreme cases oxygen and fluorine discussed above. However for the case of water it is clear that the FCH is not appropriate to describe the XAS spectra.

This work was supported by the Swedish Foundation for Strategic Research, Swedish Natural Science Research Council and National Science Foundation (US) grant CHE-0089215. Portions of this research were carried out at the Stanford Synchrotron Radiation Laboratory, a national user facility operated by Stanford University on behalf of the U.S. Department of Energy, Office of Basic Energy Sciences.

† A formal way to treat core-hole effects is to incorporate them in the two-particle Bethe-Salpeter equation[24]. Although this method dramatically improves optical-absorption calculations, in the case of XA spectra the core-hole is localized on one atomic site and the theoretical approach can be reduced to a single-particle calculation. The localization also eliminates the need for a time-dependent DFT (TDDFT) approach.

§ The DFT spectra were generated based on the oscillator strengths computed using the STOBE-DEMON program[25]. The discrete states below the edge were convoluted using gaussian functions to mimic the experimental broadening of the transitions and to ease the visual comparison with experiments; for the continuum states a larger broadening was in general applied to mimic the continuous distribution of states. For specific cluster models the broadening was done using a constant full-width-at-half-maximum (fwhm) of 0.5 eV below 537.5 eV then linearly increasing up to 8 eV at 550 eV; at higher energies the fwhm is constant. A constant broadening using gaussian functions with fwhm 0.5 eV was used for spectra computed from the MD snapshots; the final spectrum was then obtained by summing the spectra of all molecules in the snapshot. The non-local exchange functional of Becke[26] and the Perdew[27] correlation functional were used in the calculations. The core-excited oxygen was described using the IGLO-III all-electron basis set of Kutzelnigg[28], allowing for full relaxation of the core orbitals. Effective core potentials (ECP) were used on all other oxygen atoms in order to simplify the definition of the core hole[29]. The HCH and FCH methods were compared by correcting the spectral energy scale through a ΔKohn-Sham approach[18,20,30], where the XA spectrum is uniformly shifted by matching the lowest oscillator-strength to the energy difference between the total Kohn-Sham energies of the variationally obtained first resonant core-excited state and the ground state cluster. The revised value of +0.33 eV[31] for the relativistic correction for the oxygen atom is applied in the present work to all theoretical spectra.